\documentclass[twocolumn,aps,superscriptaddress,showpacs,prl,citeautoscript,footinbib]{revtex4-1}
\usepackage {natbib}
\usepackage{amsmath}
\usepackage{graphicx}
 \usepackage{color}

\begin{document}

\title{Electronic reconstruction at the isopolar LaTiO$_3$/LaFeO$_3$ interface: An x-ray photoemission and density functional theory study}

\author{J. E. Kleibeuker}
\email{jek46@cam.ac.uk}
\altaffiliation[Current address: ] {Department of Materials Science and Metallurgy, University of Cambridge, Cambridge CB3 0FS, United Kingdom}
\affiliation{Faculty of Science and Technology and MESA+ Institute for Nanotechnology, University of Twente, 7500 AE Enschede, The Netherlands}
\affiliation{Physikalisches Institut, University of W\"urzburg, 97074 W\"urzburg, Germany}
\author{Z. Zhong}
\affiliation{Institute of Solid State Physics, Vienna University of Technology, A-1040 Vienna, Austria}
\author{H. Nishikawa}
\affiliation{Faculty of Biology-Oriented Science and Technology, Kinki University, Kinokawa 649-6493, Japan}
\author{J. Gabel}
\affiliation{Physikalisches Institut, University of W\"urzburg, 97074 W\"urzburg, Germany}
\author{A. M\"uller}
\affiliation{Physikalisches Institut, University of W\"urzburg, 97074 W\"urzburg, Germany}
\author{F. Pfaff}
\affiliation{Physikalisches Institut, University of W\"urzburg, 97074 W\"urzburg, Germany}
\author{M. Sing}
\affiliation{Physikalisches Institut, University of W\"urzburg, 97074 W\"urzburg, Germany}
\author{K. Held}
\affiliation{Institute of Solid State Physics, Vienna University of Technology, A-1040 Vienna, Austria}
\author{R. Claessen}
\affiliation{Physikalisches Institut, University of W\"urzburg, 97074 W\"urzburg, Germany}
\author{G. Koster}
\affiliation{Faculty of Science and Technology and MESA+ Institute for Nanotechnology, University of Twente, 7500 AE Enschede,
The Netherlands}
\author{G. Rijnders}
\affiliation{Faculty of Science and Technology and MESA+ Institute for Nanotechnology, University of Twente, 7500 AE Enschede, The Netherlands}


\begin{abstract}
We report the formation of a non-magnetic band insulator at the isopolar interface between the antiferromagnetic Mott-Hubbard insulator LaTiO$_3$ and the antiferromagnetic charge transfer insulator LaFeO$_3$. By density functional theory calculations, we find that the formation of this interface state is driven by the combination of O~band alignment and crystal field splitting energy of the \textit{t$_{2g}$} and \textit{e$_g$} bands. As a result of these two driving forces, the Fe 3\textit{d} bands rearrange and electrons are transferred from Ti to Fe. This picture is supported by x-ray photoelectron spectroscopy, which confirms the rearrangement of the Fe~3\textit{d} bands and reveals an unprecedented charge transfer up to 1.2$\pm$0.2~\textit{e}$^-$/interface unit cell in our LaTiO$_3$/LaFeO$_3$ heterostructures. 

\end{abstract}


\maketitle

Complex oxide heterointerfaces exhibit unique properties which are absent in the corresponding isolated parent compounds \cite{Ueda1999,Gozar2008,ohtomo2004}. For example, metallic interfaces have been achieved between a polar and a non-polar insulating perovskite oxide (\textit{AB}O$_3$), e.g. at LaAlO$_3$/SrTiO$_3$, LaTiO$_3$/SrTiO$_3$ and GdTiO$_3$/SrTiO$_3$ interfaces \cite{ohtomo2004,ohtomo2002_2, Moetakef2011}. To clarify this metallic behavior, intrinsic electronic reconstruction is suggested to compensate the interfacial polar discontinuity, resulting in a quasi two dimensional electron gas at the heterointerface \cite{Okamoto2004, Nakagawa2006, Noguera2000}. However, competing mechanisms have often been proposed to act and obscure the sought-after electronic reconstruction. For example, the formation of oxygen vacancies has been shown to play an important role in the titanate-based metallic interfacial systems \cite{kalabukhov2007,siemons2007,Zhong2010, Chen2011_2}. To achieve full understanding of charge transfer, it is necessary to investigate a perovskite interface where distinct phenomena allow us to unequivocally identify the proposed charge transfer mechanism. A perovskite heterostructure where defects play no role in the physical properties is desired. Subsequently, the achieved knowledge on charge transfer in this model system can be extended to other perovskite interface systems.

In this Letter, we therefore focus on internal charge transfer at the isopolar insulating interface between LaTiO$_3$ and LaFeO$_3$, where LaTiO$_3$ is a Mott-Hubbard insulator (MHI) and LaFeO$_3$ is a charge transfer insulator (CTI) \cite{zaanen1985}. The advantage of this heterostructure is the absence of polar discontinuity at the interface. In addition, both bulk LaFeO$_3$ and bulk LaTiO$_3$ have a partially filled 3\textit{d} transition metal ion on the \textit{B}-site. This offers the opportunity to exploit the differences in band configuration of LaTiO$_3$ and LaFeO$_3$ near the Fermi level to drive electronic reconstruction.

For LaFeO$_3$, the charge transfer gap ($\Delta$) is determined by the filled oxygen 2\textit{p} band and the unoccupied upper Hubbard 3\textit{d} band of Fe ($\Delta$$_{CT}$=2.2~\textit{e}V) \cite{zaanen1985, Arima1993}. For LaTiO$_3$, the gap originates from the Mott-Hubbard splitting of the Ti \textit{d}-bands ($\Delta$$_{MH}$=0.2~\textit{e}V), while the oxygen 2\textit{p} band is located below the partially filled \textit{d}~band ($\Delta$$_{CT}$=4.5~\textit{e}V) \cite{zaanen1985, Arima1993}. In LaTiO$_3$/LaFeO$_3$ heterostructures, alignment of the O bands is expected to occur at the interface, as the two materials share their oxygen atoms at the interface \cite{Chen2013}. As a result of this band alignment, the empty upper \textit{d}~band of LaFeO$_3$ is expected to be pushed below the energy level of the partially filled lower \textit{d}~band of LaTiO$_3$, which would favor electron transfer from Ti to Fe, i.e. interfacial electronic reconstruction. 
Let us note that a charge transfer in 1:1 LaNiO$_3$/LaTiO$_3$ (CTI/MHI) superlattices has recently been studied by Chen \textit{et al.}, using density functional theory (DFT)+\textit{U} \cite{Chen2013}. The authors found that a charge transfer from Ti to Ni enhances correlation effects and leads to a Mott insulator with an enhanced moment of $S=1$ on the Ni sites and a charge transfer gap between Ni and (empty) Ti~\textit{d} states .

Based on our DFT calculations, we present clear evidence that, besides the presence of oxygen band alignment, the competition with crystal field and correlation energy of the \textit{d} electrons is crucial to achieve electronic reconstruction at MHI/CTI interfaces. At LaTiO$_3$/LaFeO$_3$ interfaces, this competition results in both charge transfer and a rearrangement of the Fe bands which can lead to a new non-magnetic band insulating state at the interface. Using \textit{in situ} X-ray photoelectron spectroscopy (XPS), we confirm the charge transfer and band rearrangement experimentally. By fitting the XPS data, we have determined an electron transfer up to 1.2$\pm$0.2 per interface unit cell (u.c.) from Ti to Fe. 

\begin{figure}
\begin{center}
\includegraphics[width=0.46\textwidth]{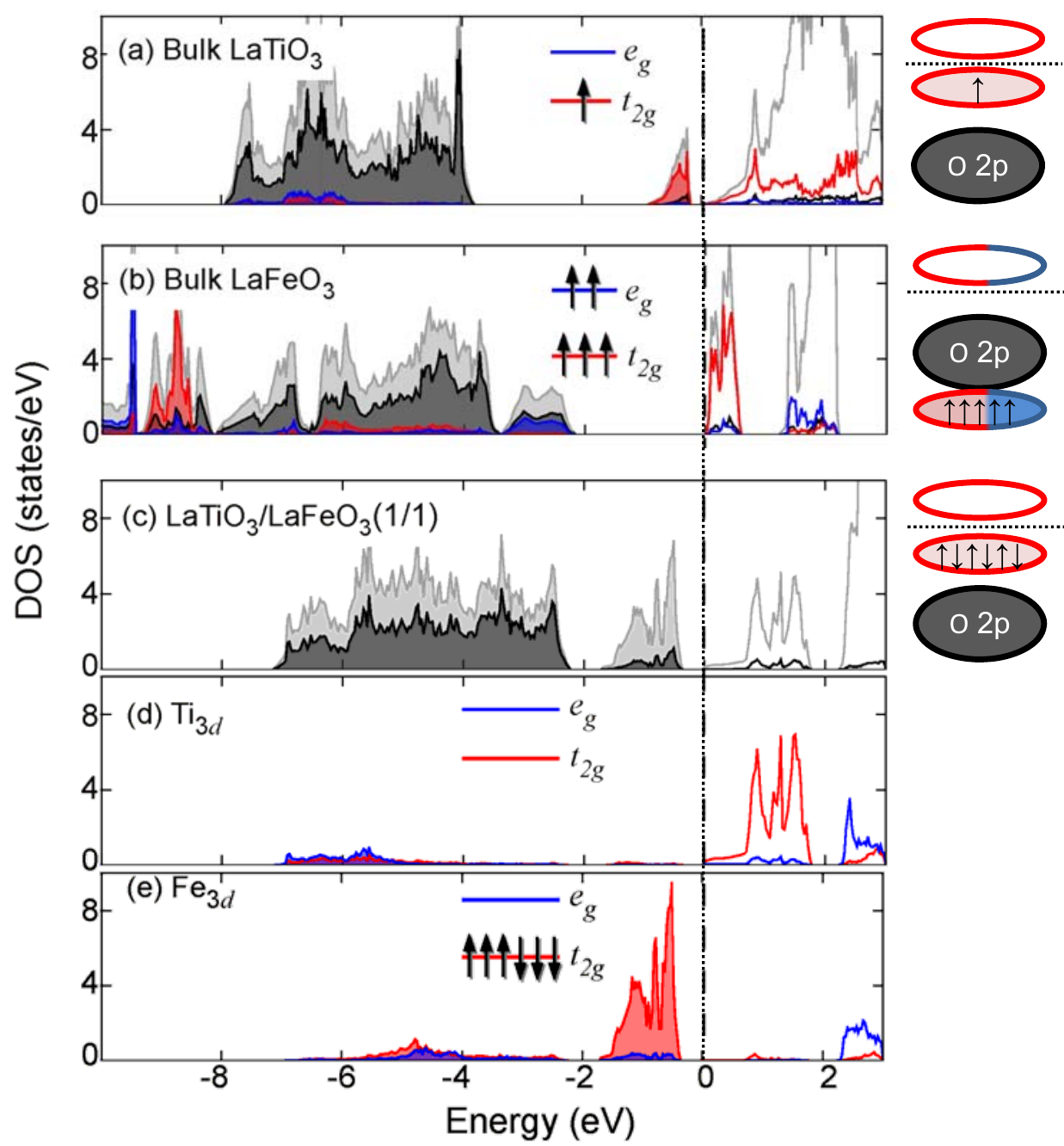}
\caption{\label{DOS}Atomic and orbital projected DOS as well as schematic band structure of (a) bulk LaTiO$_3$, (b) bulk LaFeO$_3$, and (c-e) a (1/1) LaTiO$_3$/LaFeO$_3$ superlattice. Total states are marked in grey; O ${p}$ states in black; Fe and Ti $t_{2g}$ states in red, and Fe and Ti $e_g$ states in blue. The Fermi level is indicated by the dotted line. }
\end{center}
\end{figure}

For the DFT calculations, we employed the local density approximation (LDA) and the projector augmented-wave method as implemented in the Vienna \textit{ab-initio} simulation package (\footnotesize{VASP}\normalsize) \cite{Blochl1994,Kresse1999}. A kinetic energy cutoff of 500 \textit{e}V was used and the Brillouin zone was sampled with an 8$\times$8$\times$6 \textit{k}-point grid in combination with a tetrahedron method. Including an on-site Coulomb interaction, the LDA+U calculated ground states and energy gaps for bulk LaTiO$_3$ and LaFeO$_3$ agree well with experiments for an optimized U$_\textit{d}^{Ti}$=3.0~\textit{e}V and U$_\textit{d}^{Fe}$=4.8~\textit{e}V, respectively (see Figs.~\ref{DOS}a and \ref{DOS}b) \cite{Dudarev1998, tokura1993, Koehler1960}. Bulk LaTiO$_3$ had a MHI-type energy gap between the filled and unfilled Ti $t_{2g}$ states and bulk LaFeO$_3$ had a CTI-type energy gap between the filled O $2p$ states hybridized to Fe $e_g$ states and the unfilled Fe $t_{2g}$ states~\cite{Pavarini2004}. Both bulk materials were G-type antiferromagnetic. Subsequently, we modeled (1/1), (2/2) and (2/4) LaTiO$_3$/LaFeO$_3$ heterostructures using a periodically repeated supercell \cite{Zhong2008}. The unit cells had a GdFeO$_3$-type distorted orthorhombic structure and the lattice constants were fixed at the optimized LaTiO$_3$ bulk values~\cite{Pavarini2004}. The atoms were allowed to relax internally. To integrate these distortions in LaTiO$_3$/LaFeO$_3$ superlattices, we replaced one Ti atom of the distorted LaTiO$_3$ structure, which has a $\sqrt{2}$\textit{a$_{pc}$}$\times$$\sqrt{2}$\textit{a$_{pc}$}$\times$2\textit{c$_{pc}$} structure, by an Fe atom along the \textit{c}-axis. 

The atomic and orbital projected density of states (DOS) of a (1/1) LaTiO$_3$/LaFeO$_3$ superlattice are shown in Figs.~\ref{DOS}c-e. At the interface, the non-bonding oxygen bands of LaTiO$_3$ and LaFeO$_3$ align (Fig.~\ref{DOS}c), the Ti 3\textit{d} bands are empty (Fig.~\ref{DOS}d) and 6 electrons are located in the Fe~3\textit{d} band (Fig.~\ref{DOS}e). This means that one electrons is transferred from Ti to Fe, resulting in Ti$^{4+}$ and Fe$^{2+}$. In addition, a rearrangement of the Fe~3$d$ bands in the LaTiO$_3$/LaFeO$_3$ superlattice is observed. Here, a completely filled Fe~\textit{t$_{2g}$} band is located above the O~2\textit{p} band and the Fe \textit{e$_g$} band is empty (Fig.~\ref{DOS}e), while in bulk the filled lower Hubbard band of Fe is below the O~2\textit{p} band (Fig.~\ref{DOS}b). Due to the electron transfer and band rearrangement, a band insulator (BI) with a gap between the filled Fe~\textit{t$_{2g}$} and the empty Ti~\textit{t$_{2g}$} bands ($\Delta$$_{B}$$\approx$0.5~\textit{e}V) is formed at the interface \cite{zaanen1985}. In addition, the DFT results point to a magnetic transition: from Ti$^{3+}$($t_{2g}$) and high spin Fe$^{3+}$ (3$t_{2g}$$\uparrow$, 2$e_{g}$$\uparrow$) configuration in bulk to Ti$^{4+}$ and low spin Fe$^{2+}$ (3$t_{2g}$$\uparrow$, 3$t_{2g}$$\downarrow$) configuration (i.e. non-magnetic) at the interface. To ensure that the observed charge transfer depended on the presence of partially filled \textit{d}~bands on both sides of the interface, we also calculated (1/1) and (2/2) LaAlO$_3$/LaFeO$_3$ superlattices. Here, no electron transfer or magnetic transition occurs, since Al has an empty 3\textit{d}~band well above the Fermi energy, which fixes the Al valence strictly to 3+ (see also Fig. 1 of Supplemental Material)~\cite{SupplMat}.

According to the DFT results, the observed charge transfer at the LaTiO$_3$/LaFeO$_3$ interface is very robust. Increasing the thickness of LaFeO$_3$ to 4 u.c., slightly straining of the unit cells, or varying U$_{Ti, Fe}$ between 0 and 5 \textit{e}V does not eliminate the observed transfer of one electron per interface unit cell. Moreover, investigating a (2/4) LaTiO$_3$/LaFeO$_3$ superlattice, it appears that the majority of transferred electron remains at the LaFeO$_3$ interface layer (Fig.2c-e of Supplemental Material~\cite{SupplMat}). The layers further away from the interface, closely resemble the bulk DOS of LaFeO$_3$ (Fig.~\ref{DOS}b). Let us note that the interface charge transfer is very robust and reliable for any LaFeO$_3$ thickness. Even LaTiO$_3$/LaFeO$_3$ heterostructures without structural distortions show this one electron charge transfer (See Supplemental Material~\cite{SupplMat}). Since the charge transfer may lead to complex physical behavior in LaFeO$_3$, as a result of the competition of various magnetic configurations (bulk vs. interface), it is difficult to accurately determine the magnetic and electronic state of interfaces where LaFeO$_3$ $>$~2 u.c.

The DFT results indicated that the interfacial electron transfer at LaTiO$_3$/LaFeO$_3$ interfaces is the consequence of (i) electrochemical potential, also described as O band alignment, and (ii) crystal field splitting and Hund's exchange. Taking only the O band alignment into account, electrons flow from Ti to Fe and reduce their electrochemical potential. As a result, an internal electric field, which balances the electrochemical potential difference between Ti and Fe, is created and prevents further charge transfer. This is also the reason why charge transfer at oxide interfaces is not evident when it only relies on O band alignment \cite{Zubko2011}. In LaTiO$_3$/LaFeO$_3$, however, an additional force comes into play, namely a rearrangement of the Fe~3\textit{d} bands. The origin of this rearrangement is a high-spin to low-spin transition which is a result of the competition between Hund's exchange and crystal field splitting (see Supplemental Material~\cite{SupplMat}). This makes the low-spin configuration energetically more favorable for Fe$^{2+}$ and yields an additional energy gain for the charge transfer. As a result, a strong electron transfer is observed at the LaTiO$_3$/LaFeO$_3$ interface and accompanied by a loss of magnetic moment.

\begin{figure}[t]
\begin{center}
\includegraphics[width=0.48\textwidth]{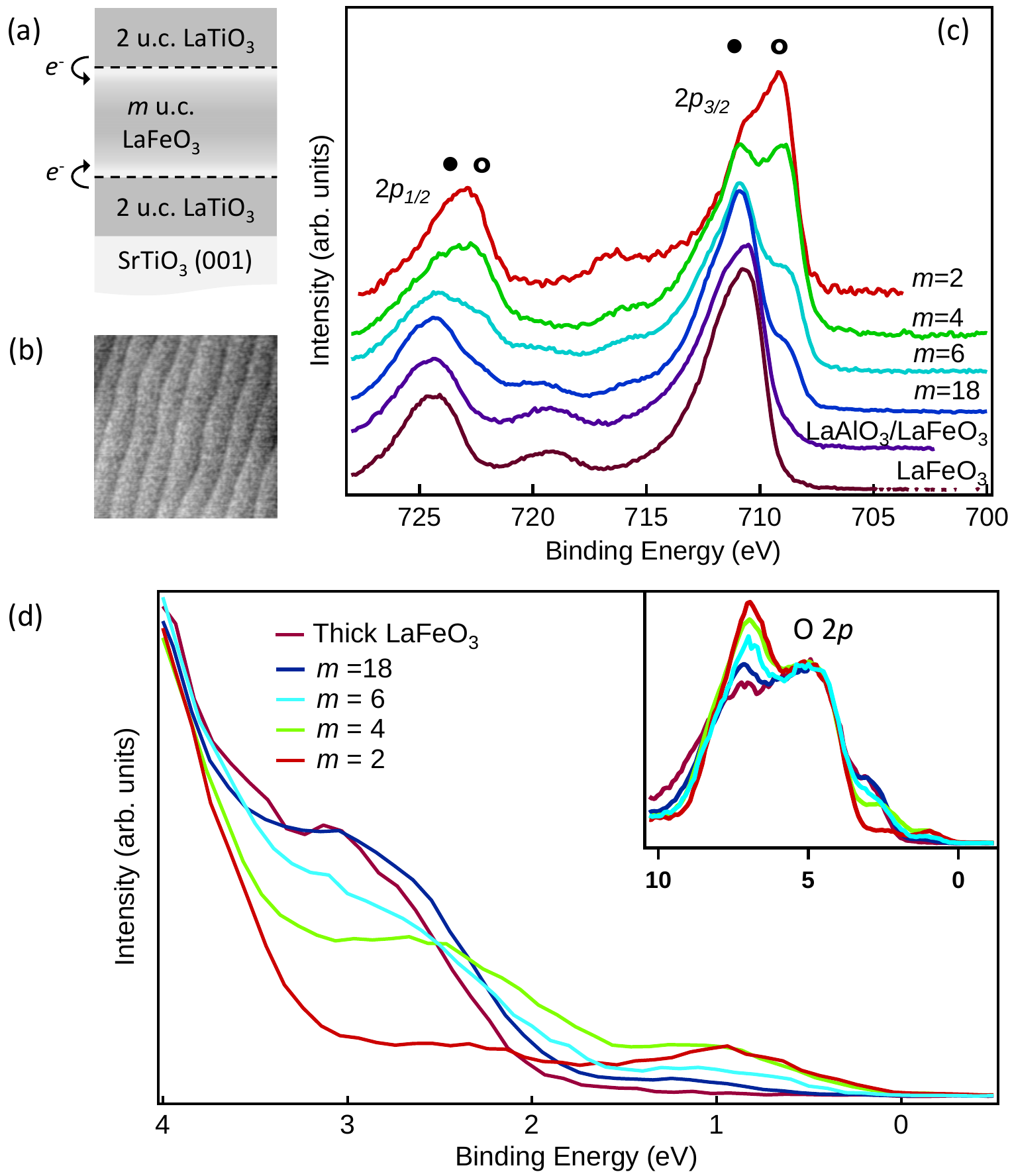}
\caption{\label{Febasic} (a) Sketch of the LaTiO$_3$/LaFeO$_3$ sample geometry. (b) A typical 1$\times$1 $\mu$m AFM height image of a LaTiO$_3$/LaFeO$_3$ heterostructure. (c) Fe 2\textit{p} XPS spectra of LaTiO$_3$/LaFeO$_3$ heterostructures for various thicknesses of LaFeO$_3$, as well as of a 30 u.c. LaFeO$_3$ film and a (2/2) LaAlO$_3$/LaFeO$_3$ heterostructure. The solid and open circles mark the Fe$^{3+}$ and Fe$^{2+}$ peaks respectively. (d) Valence band XPS spectra of LaTiO$_3$/LaFeO$_3$ heterostructures for various thicknesses of LaFeO$_3$. All spectra were taken near normal emission ($\theta$=3$^\circ$). } 
\end{center}
\end{figure}

To resolve the predicted charge transfer and band rearrangement experimentally, we used XPS. XPS is very sensitive to variations in the valence state of transition metal ions and able to detect the valence band structure. Therefore, it is a perfectly suited technique to determine the presence of both charge transfer and band rearrangement at the LaTiO$_3$/LaFeO$_3$ interface. We have studied LaTiO$_3$/LaFeO$_3$ heterostructures where the LaFeO$_3$ layer (\textit{m} = 2, 4, 6, 18 u.c.) was sandwiched between two LaTiO$_3$ layers, each 2 u.c. thick (see Fig.~\ref{Febasic}a). The heterostructures were grown on TiO$_2$-terminated SrTiO$_3$ (001) single crystals using pulsed laser deposition \cite{koster1998}. Commercial LaFeO$_3$ and La$_2$Ti$_2$O$_7$ sintered targets were ablated at a fluence of 1.9 Jcm$^{-2}$ and a repetition rate of 1 Hz. During growth, the substrate was held at 730~$^\circ$C in 2$\times$10$^{-6}$ mbar oxygen atmosphere. Subsequently, the samples were cooled down to room temperature in 2$\times$10$^{-6}$ mbar oxygen. The low growth pressure was chosen to ensure the fabrication of the perovskite phase of LaTiO$_3$ \cite{ohtomo2002}.

The growth was \textit{in situ} monitored by reflection high-energy electron diffraction (RHEED). Clear oscillations were observed during deposition and the RHEED pattern remained two dimensional \footnote{Note that some Ti/Fe intermixing across the interface may be present, taking the low oxygen pressure during growth into account~\cite{Willmott2007}.}. Atomically smooth film surfaces with a defined terrace structure and one unit cell steps ($\sim$0.4~nm) were confirmed by atomic force microscopy (AFM) (see Fig.~\ref{Febasic}b). X-ray diffraction reciprocal space maps showed that the heterostructures were fully strained and that the LaTiO$_3$ and LaFeO$_3$ u.c. volumes were similar to their bulk values. The volume conservation indicates that the heterostructures had a low defect density. The possible conducting behavior of the heterointerfaces could not be verified since the transport measurements were dominated by oxygen deficient SrTiO$_3$ as a result of the low pressure during growth and cool down.

Directly after growth, the LaTiO$_3$/LaFeO$_3$ heterostructures were measured by \textit{in situ} XPS (see Fig~\ref{Febasic}c and~\ref{Febasic}d). The XPS system was equipped with an EA 125 electron energy analyzer. The measurements were done using a monochromized Al K$\alpha$ source (1486.6~\textit{e}V). All spectra were aligned to the O~1\textit{s} at 530.1~\textit{e}V \footnote{No charging of the samples was observed during X-ray exposure since the SrTiO$_{3-\delta}$ became conducting as a result of the low oxygen pressure during growth and cool down.}. For analysis of the Fe~2\textit{p} spectra, a Shirley background was subtracted and the spectra were normalized to the total area \footnote{The La MNN (at $\sim$740-800~\textit{e}V) obscures the Fe 2\textit{p} satellite structure at higher binding energy. To allow proper normalization, we limited the Fe~2\textit{p} range up to this satellite peak.}. The valence band spectra were normalized to the intensity of the O~2\textit{p} peak at 5~\textit{e}V \footnote{\label{note1}Normalization of the valence band spectra is complicated by the Ti-O~2\textit{p} and Fe-O~2\textit{p} hybridization. To allow for a qualitative analysis, the valence band spectra were aligned on the intensity of the O~2\textit{p} at 5~\textit{e}V. However, this may result in minor normalization artefacts.}.  

Fig.~\ref{Febasic}c shows the Fe 2\textit{p} spectra of LaTiO$_3$/LaFeO$_3$ heterostructures and a 30 u.c. thick LaFeO$_3$ film. The LaFeO$_3$ film exhibits a typical Fe$^{3+}$ spectrum \cite{Fujii1999}. For the LaTiO$_3$/LaFeO$_3$ heterostructures, additional spectral weight is present at $\sim$2~\textit{e}V lower binding energy. This suggests that both Fe$^{3+}$ and Fe$^{2+}$ are present in the heterostructures and indicates that Fe reduction occurs adjacent to LaTiO$_3$. For comparison, only Fe$^{3+}$ is observed in LaFeO$_3$ (\textit{m}=2) sandwiched between LaAlO$_3$ layers (Fig~\ref{Febasic}c). Reducing the thickness of the LaFeO$_3$ layer in the heterostructures resulted in an increase of the Fe$^{2+}$ signal, which confirms the DFT prediction that electron transfer occurs at LaTiO$_3$/LaFeO$_3$ interfaces.  We also measured the Ti~2\textit{p} spectra of the heterostructures to determine the presence of both Ti$^{3+}$ to Ti$^{4+}$. Here, however, only a single peak for both the Ti~2\textit{p}$_{3/2}$ (at 459 eV) and Ti~2\textit{p}$_{1/2}$ spin-orbit peaks is observed. This could indicate a single Ti valence of presumably 4+ and hence complete charge transfer from Ti to Fe across the interface, independent of LaFeO$_3$ thickness in agreement with our DFT+U calculations (see Supplemental Material~\cite{SupplMat})~\cite{Kareev2013}.

\begin{figure}[t]
\begin{center}
\includegraphics[width=0.43\textwidth]{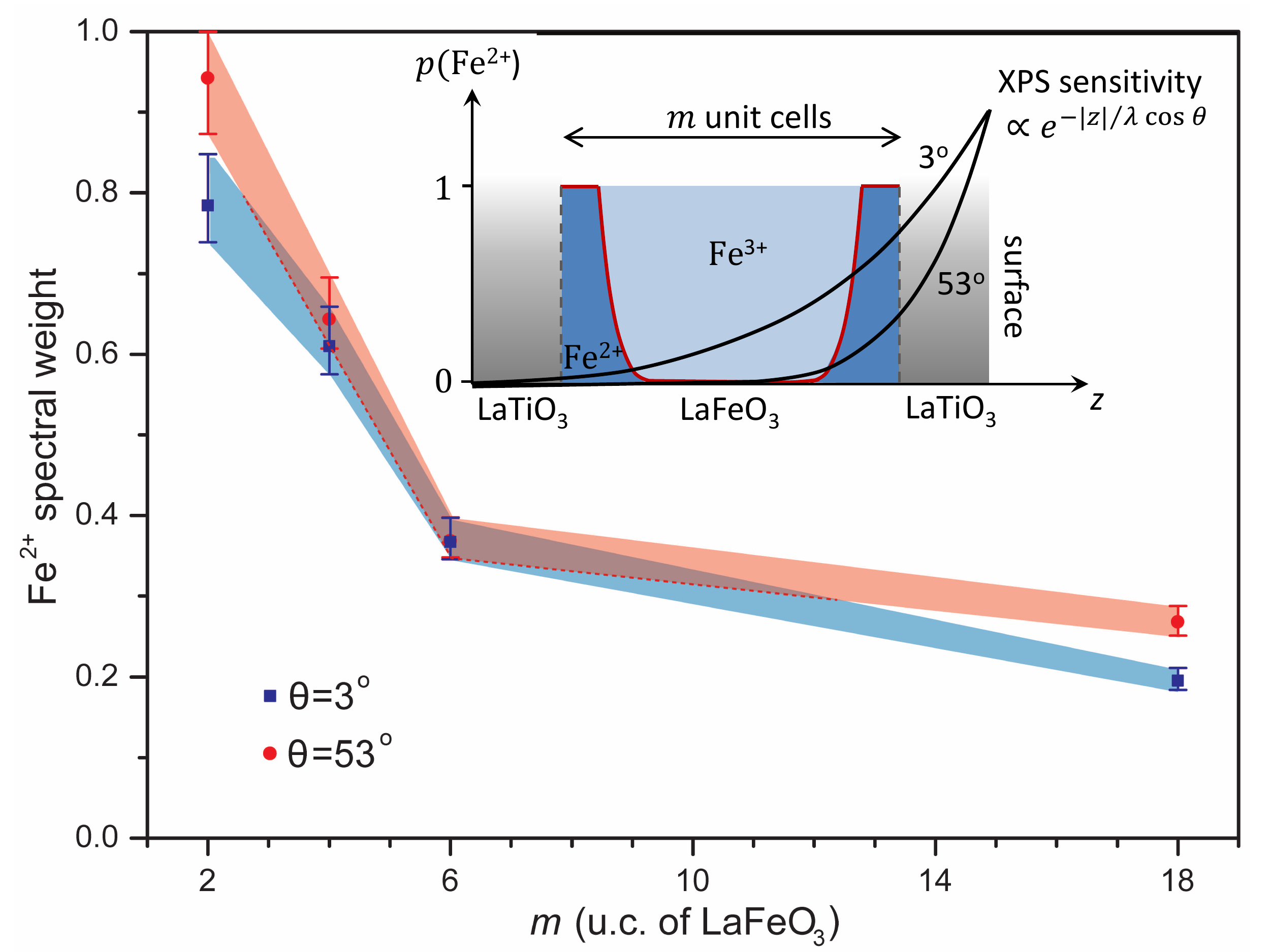}
\caption{\label{Fespectrafit} Fe$^{2+}$ spectral weight versus LaFeO$_3$ thickness for both bulk (blue) and surface  (red) sensitive XPS measurements, taking [Fe$_{total}$]=[Fe$^{3+}$]+[Fe$^{2+}$]=1. The  inset  is  a schematic view of the Fe$^{2+}$ fraction (\textit{p}(Fe$^{2+}$)) across the LaFeO$_3$ layer (indicated by the solid red curve). Fe$^{2+}$ (Fe$^{3+}$) fraction is given in dark (light) blue. In addition, an indication of the XPS sensitivity for both surface (53$^\circ$) and bulk (3$^\circ$) sensitive measurements is shown. \textit{z} indicates the direction perpendicular to the surface. }
\end{center}
\end{figure}

To quantify the total number of electrons transferred from LaTiO$_3$ to LaFeO$_3$ as well as the electron distribution across the LaFeO$_3$ layer, we performed angular resolved XPS measurements. By varying the emission angle $\theta$ with respect to the surface normal, we controlled the probing depth, i.e. controlled the effective electron escape depth $\lambda_{eff}$=$\lambda\cos\theta$, where $\lambda$ is approximately 1.7~nm (see inset Fig.~\ref{Fespectrafit})~\cite{NIST}. Next, we determined the Fe$^{2+}$ and Fe$^{3+}$ fractions of the Fe~2\textit{p} spectra by decomposing the Shirley corrected spectra into an Fe$^{2+}$ and Fe$^{3+}$ component (see for more details Supplemental Material~\cite{SupplMat}). This resulted in a window of Fe$^{2+}$ XPS signal for bulk ($\theta$=3$^\circ$) and surface ($\theta$=53$^\circ$) sensitive measurements, which is shown in Fig.~\ref{Fespectrafit}. Both the decrease in spectral weight of Fe$^{2+}$ for increasing LaFeO$_3$ thickness and the stronger Fe$^{2+}$ signal in the surface sensitive measurements suggest that the transferred electrons are located near the LaTiO$_3$/LaFeO$_3$ interface. Note that the difference between the bulk and surface sensitive measurement for the \textit{m}=2 LaTiO$_3$/LaFeO$_3$ heterostructure would not be present if both LaTiO$_3$/LaFeO$_3$ interfaces behaved equally. For this specific sample, however, the deposition length of the top LaTiO$_3$ layer was 7\% (2 pulses) longer than for the bottom LaTiO$_3$ layer. This may explain the difference between the bulk and surface sensitive measurements. In addition, the underlying SrTiO$_3$/LaTiO$_3$ interface may also reduce the total electron transfer from the bottom LaTiO$_3$ layer to the LaFeO$_3$ layer~\cite{ohtomo2002_2}.

Subsequently, we determined the total electron transfer and electron distribution by modelling the thickness dependence of the spectral weight of Fe$^{2+}$ shown in Fig.~\ref{Fespectrafit}. This was done by iteratively optimizing the electron doping in the five LaFeO$_3$ layers nearest to the interface with LaTiO$_3$ between 0 and 1 (for more details see Supplemental Material~\cite{SupplMat}). This model confirmed that the majority of transferred electrons was located in the LaFeO$_3$ layer closest to the interface as well as that the number of electrons rapidly decreased for layers further away from the interface (see inset Fig.~\ref{Fespectrafit}). These findings are in good agreement with our DFT results, where for thicker LaFeO$_3$ layers also a minor part of the electrons is transferred to the LaFeO$_3$ layers away from the interface (see Supplemental Material Fig.~2e~\cite{SupplMat}). In addition, the model gave an indication of the total electron transfer, from 0.8$-$1.0 \textit{e}$^-$/interface~u.c. for \textit{m}~=~2 heterostructures to 1.1$-$1.4 \textit{e}$^-$/interface~u.c. for heterostructures with \textit{m}~$>$~10. The total electron transfer being $>$~1~\textit{e}$^-$/interface~u.c. indicates that additional electrons are transferred from the LaTiO$_3$ layers further away from the interface. This is also suggested by our DFT results taking Ti surface states into account (see Supplemental Material~\cite{SupplMat}). In comparison to our DFT results, the total charge transfer observed experimentally is significantly higher. However, for the DFT calculations a (1/1) system was used, thus all LaTiO$_3$ layers being adjacent to LaFeO$_3$, and therefore, the number of transferred electrons could not exceed 1~\textit{e}$^-$/interface~u.c. Let us note that possible Ti/Fe intermixing across the interface may affect the exact electron distribution and total charge transfer, but does not change the essential interface physics (see also Supplemental Material~\cite{SupplMat}).

Next to electron transfer, our DFT calculations predict rearrangement of the Fe~3\textit{d} bands. To study this rearrangement, we measured the valence band spectra by XPS (Fig.~\ref{Febasic}d) \footnotemark[4]. 
Comparing the spectra of LaTiO$_3$/LaFeO$_3$ heterostructures with the spectra of the thick LaFeO$_3$ film, a new peak at 1~\textit{e}V is present for the heterostructures. According to the DFT calculations, this new peak is attributed to the completely filled \textit{t$_{2g}$} band of Fe$^{2+}$. The intensity of this peak depends on the number of  strongly electron doped LaFeO$_3$ layers near the surface. Taking the electron distribution in LaFeO$_3$ into account, the first two LaFeO$_3$ layers nearest to the LaTiO$_3$/LaFeO$_3$ interface would mainly contribute to the spectral weight of this peak. This also explains the similar peak intensity for the \textit{m}=2 and \textit{m}=4 heterostructures, but reduced intensity for the thicker heterostructures. Simultaneously, the charge transfer band of LaFeO$_3$, resulting from the O~2\textit{p}-Fe~\textit{e$_g$} hybridization, decreases in intensity.  This strongly supports the occurrence of Fe band rearrangement at the LaTiO$_3$/LaFeO$_3$ interface predicted by DFT. The presence of Fe band rearrangement strongly indicates that the interfaces become non-magnetic, as proposed by our DFT calculations. In addition, the Ti 3\textit{d}$^1$ band near the Fermi level may be present in the valence band spectra. However, the resulting changes in the Ti~3\textit{d} occupation of the LaTiO$_3$ layers are difficult to extract from the spectra shown in Fig.~\ref{Febasic}d, as the Ti 3\textit{d}$^1$ peak is very weak and probably obscured by the appearance of the new Fe peak \cite{Takizawa2006}.

In conclusion, we have shown that the competition between electrochemical potential, crystal field splitting and correlation energy can lead to an unprecedented transfer of electrons at LaTiO$_3$/LaFeO$_3$ interfaces. Using XPS, we showed a charge transfer up to 1.2$\pm$0.2~\textit{e}$^-$/interface~u.c. from Ti to Fe as well as the rearrangement of the Fe~3\textit{d} bands. For LaTiO$_3$/LaFeO$_3$, the charge transfer suppresses the magnetic moment and antiferromagnetism at the interface.  Considering the basic electronic configuration, we expect however the interfaces of e.g. LaTiO$_3$/LaMnO$_3$ and LaTiO$_3$/LaCoO$_3$ to become ferromagnetic upon charge transfer. Moreover, by applying biaxial strain, it may be possible to control the number of transferred electrons and, with it, the interfacial properties. Hence, the reported charge transfer up to 1.2$\pm$0.2~\textit{e}$^-$/interface u.c. opens novel routes to design functional oxide interfaces.

The authors thank B. Kuiper for valuable technical help. G.R. thanks the financial support by The Netherlands Organization for Scientific Research (NWO) through a VIDI grant. R.C. and K.H. acknowledge support from Research Unit FOR 1346 of the Deutsche Forschungsgemeinschaft and the Austrian Science Fund (project ID  I597), respectively.


\begin{thebibliography}{34}%
\makeatletter
\providecommand \@ifxundefined [1]{%
 \@ifx{#1\undefined}
}%
\providecommand \@ifnum [1]{%
 \ifnum #1\expandafter \@firstoftwo
 \else \expandafter \@secondoftwo
 \fi
}%
\providecommand \@ifx [1]{%
 \ifx #1\expandafter \@firstoftwo
 \else \expandafter \@secondoftwo
 \fi
}%
\providecommand \natexlab [1]{#1}%
\providecommand \enquote  [1]{``#1''}%
\providecommand \bibnamefont  [1]{#1}%
\providecommand \bibfnamefont [1]{#1}%
\providecommand \citenamefont [1]{#1}%
\providecommand \href@noop [0]{\@secondoftwo}%
\providecommand \href [0]{\begingroup \@sanitize@url \@href}%
\providecommand \@href[1]{\@@startlink{#1}\@@href}%
\providecommand \@@href[1]{\endgroup#1\@@endlink}%
\providecommand \@sanitize@url [0]{\catcode `\\12\catcode `\$12\catcode
  `\&12\catcode `\#12\catcode `\^12\catcode `\_12\catcode `\%12\relax}%
\providecommand \@@startlink[1]{}%
\providecommand \@@endlink[0]{}%
\providecommand \url  [0]{\begingroup\@sanitize@url \@url }%
\providecommand \@url [1]{\endgroup\@href {#1}{\urlprefix }}%
\providecommand \urlprefix  [0]{URL }%
\providecommand \Eprint [0]{\href }%
\providecommand \doibase [0]{http://dx.doi.org/}%
\providecommand \selectlanguage [0]{\@gobble}%
\providecommand \bibinfo  [0]{\@secondoftwo}%
\providecommand \bibfield  [0]{\@secondoftwo}%
\providecommand \translation [1]{[#1]}%
\providecommand \BibitemOpen [0]{}%
\providecommand \bibitemStop [0]{}%
\providecommand \bibitemNoStop [0]{.\EOS\space}%
\providecommand \EOS [0]{\spacefactor3000\relax}%
\providecommand \BibitemShut  [1]{\csname bibitem#1\endcsname}%
\let\auto@bib@innerbib\@empty
\bibitem [{\citenamefont {Ueda}\ \emph {et~al.}(1999)\citenamefont {Ueda},
  \citenamefont {Tabata},\ and\ \citenamefont {Kawai}}]{Ueda1999}%
  \BibitemOpen
  \bibfield  {author} {\bibinfo {author} {\bibfnamefont {K.}~\bibnamefont
  {Ueda}}, \bibinfo {author} {\bibfnamefont {H.}~\bibnamefont {Tabata}}, \ and\
  \bibinfo {author} {\bibfnamefont {T.}~\bibnamefont {Kawai}},\ }\href@noop {}
  {\bibfield  {journal} {\bibinfo  {journal} {Phys. Rev. B}\ }\textbf {\bibinfo
  {volume} {60}},\ \bibinfo {pages} {R12561} (\bibinfo {year} {1999})}\BibitemShut
  {NoStop}%
\bibitem [{\citenamefont {Gozar}\ \emph {et~al.}(2008)\citenamefont {Gozar},
  \citenamefont {Logvenov}, \citenamefont {Fitting~Kourkoutis}, \citenamefont
  {Bollinger}, \citenamefont {Giannuzzi}, \citenamefont {Muller},\ and\
  \citenamefont {Bozovic}}]{Gozar2008}%
  \BibitemOpen
  \bibfield  {author} {\bibinfo {author} {\bibfnamefont {A.}~\bibnamefont
  {Gozar}}, \bibinfo {author} {\bibfnamefont {G.}~\bibnamefont {Logvenov}},
  \bibinfo {author} {\bibfnamefont {L.}\ \bibnamefont {Fitting~Kourkoutis}},
  \bibinfo {author} {\bibfnamefont {A.~T.}\ \bibnamefont {Bollinger}}, \bibinfo
  {author} {\bibfnamefont {L.~A.}\ \bibnamefont {Giannuzzi}}, \bibinfo {author}
  {\bibfnamefont {D.~A.}\ \bibnamefont {Muller}}, \ and\ \bibinfo {author}
  {\bibfnamefont {I.}~\bibnamefont {Bozovic}},\ }\href@noop {} {\bibfield
  {journal} {\bibinfo  {journal} {Nature}\ }\textbf {\bibinfo {volume} {455}},\
  \bibinfo {pages} {782} (\bibinfo {year} {2008})}\BibitemShut {NoStop}%
\bibitem [{\citenamefont {Ohtomo}\ and\ \citenamefont
  {Hwang}(2004)}]{ohtomo2004}%
  \BibitemOpen
  \bibfield  {author} {\bibinfo {author} {\bibfnamefont {A.}~\bibnamefont
  {Ohtomo}}\ and\ \bibinfo {author} {\bibfnamefont {H.~Y.}\ \bibnamefont
  {Hwang}},\ }\href@noop {} {\bibfield  {journal} {\bibinfo  {journal}
  {Nature}\ }\textbf {\bibinfo {volume} {427}},\ \bibinfo {pages} {423}
  (\bibinfo {year} {2004})}\BibitemShut {NoStop}%
\bibitem [{\citenamefont {Ohtomo}\ \emph
  {et~al.}(2002{\natexlab{a}})\citenamefont {Ohtomo}, \citenamefont {Muller},
  \citenamefont {Grazul},\ and\ \citenamefont {Hwang}}]{ohtomo2002_2}%
  \BibitemOpen
  \bibfield  {author} {\bibinfo {author} {\bibfnamefont {A.}~\bibnamefont
  {Ohtomo}}, \bibinfo {author} {\bibfnamefont {D.~A.}\ \bibnamefont {Muller}},
  \bibinfo {author} {\bibfnamefont {J.~L.}\ \bibnamefont {Grazul}}, \ and\
  \bibinfo {author} {\bibfnamefont {H.~Y.}\ \bibnamefont {Hwang}},\ }\href@noop
  {} {\bibfield  {journal} {\bibinfo  {journal} {Nature}\ }\textbf {\bibinfo
  {volume} {419}},\ \bibinfo {pages} {378} (\bibinfo {year}
  {2002}{\natexlab{a}})}\BibitemShut {NoStop}%
\bibitem [{\citenamefont {Moetakef}\ \emph {et~al.}(2011)\citenamefont
  {Moetakef}, \citenamefont {Cain}, \citenamefont {Ouellette}, \citenamefont
  {Zhang}, \citenamefont {Klenov}, \citenamefont {Janotti}, \citenamefont
  {Van~de Walle}, \citenamefont {Rajan}, \citenamefont {Allen},\ and\
  \citenamefont {Stemmer}}]{Moetakef2011}%
  \BibitemOpen
  \bibfield  {author} {\bibinfo {author} {\bibfnamefont {P.}~\bibnamefont
  {Moetakef}}, \bibinfo {author} {\bibfnamefont {T.~A.}~\bibnamefont {Cain}},
  \bibinfo {author} {\bibfnamefont {D.~G.}~\bibnamefont {Ouellette}}, \bibinfo
  {author} {\bibfnamefont {J.~Y.}\ \bibnamefont {Zhang}}, \bibinfo {author}
  {\bibfnamefont {D.~O.}\ \bibnamefont {Klenov}}, \bibinfo {author}
  {\bibfnamefont {A.}~\bibnamefont {Janotti}}, \bibinfo {author} {\bibfnamefont
  {C.~G.}\ \bibnamefont {Van~de Walle}}, \bibinfo {author} {\bibfnamefont
  {S.}~\bibnamefont {Rajan}}, \bibinfo {author} {\bibfnamefont {S.~J.}\
  \bibnamefont {Allen}}, \ and\ \bibinfo {author} {\bibfnamefont
  {S.}~\bibnamefont {Stemmer}},\ }\href@noop {} {\bibfield  {journal} {\bibinfo
   {journal} {Appl. Phys. Lett.}\ }\textbf {\bibinfo {volume} {99}},\ \bibinfo
  {pages} {232116} (\bibinfo {year} {2011})}\BibitemShut {NoStop}%
\bibitem [{\citenamefont {Okamoto}\ and\ \citenamefont
  {Millis}(2004)}]{Okamoto2004}%
  \BibitemOpen
  \bibfield  {author} {\bibinfo {author} {\bibfnamefont {S.}~\bibnamefont
  {Okamoto}}\ and\ \bibinfo {author} {\bibfnamefont {A.}~\bibnamefont
  {Millis}},\ }\href@noop {} {\bibfield  {journal} {\bibinfo  {journal}
  {Nature}\ }\textbf {\bibinfo {volume} {428}},\ \bibinfo {pages} {630}
  (\bibinfo {year} {2004})}\BibitemShut {NoStop}%
\bibitem [{\citenamefont {Nakagawa}\ \emph {et~al.}(2006)\citenamefont
  {Nakagawa}, \citenamefont {Hwang},\ and\ \citenamefont
  {Muller}}]{Nakagawa2006}%
  \BibitemOpen
  \bibfield  {author} {\bibinfo {author} {\bibfnamefont {N.}~\bibnamefont
  {Nakagawa}}, \bibinfo {author} {\bibfnamefont {H.~Y.}~\bibnamefont {Hwang}}, \
  and\ \bibinfo {author} {\bibfnamefont {D.~A.}~\bibnamefont {Muller}},\
  }\href@noop {} {\bibfield  {journal} {\bibinfo  {journal} {Nat. Mater.}\
  }\textbf {\bibinfo {volume} {5}},\ \bibinfo {pages} {204} (\bibinfo {year}
  {2006})}\BibitemShut {NoStop}%
\bibitem [{\citenamefont {Noguera}(2000)}]{Noguera2000}%
  \BibitemOpen
  \bibfield  {author} {\bibinfo {author} {\bibfnamefont {C.}~\bibnamefont
  {Noguera}},\ }\href@noop {} {\bibfield  {journal} {\bibinfo  {journal}
  {J. Phys.: Condens. Matter.}\ }\textbf {\bibinfo {volume} {12}},\ \bibinfo {pages} {R367}
  (\bibinfo {year} {2000})}\BibitemShut {NoStop}%
\bibitem [{\citenamefont {Kalabukhov}\ \emph {et~al.}(2007)\citenamefont
  {Kalabukhov}, \citenamefont {Gunnarsson}, \citenamefont {B\"orjesson},
  \citenamefont {Olsson}, \citenamefont {Claeson},\ and\ \citenamefont
  {Winkler}}]{kalabukhov2007}%
  \BibitemOpen
  \bibfield  {author} {\bibinfo {author} {\bibfnamefont {A.}~\bibnamefont
  {Kalabukhov}}, \bibinfo {author} {\bibfnamefont {R.}~\bibnamefont
  {Gunnarsson}}, \bibinfo {author} {\bibfnamefont {J.}~\bibnamefont
  {B\"orjesson}}, \bibinfo {author} {\bibfnamefont {E.}~\bibnamefont {Olsson}},
  \bibinfo {author} {\bibfnamefont {T.}~\bibnamefont {Claeson}}, \ and\
  \bibinfo {author} {\bibfnamefont {D.}~\bibnamefont {Winkler}},\ }\href@noop
  {} {\bibfield  {journal} {\bibinfo  {journal} {Phys. Rev. B}\ }\textbf
  {\bibinfo {volume} {75}},\ \bibinfo {pages} {121404} (\bibinfo {year}
  {2007})}\BibitemShut {NoStop}%
\bibitem [{\citenamefont {Siemons}\ \emph {et~al.}(2007)\citenamefont
  {Siemons}, \citenamefont {Koster}, \citenamefont {Yamamoto}, \citenamefont
  {Harrison}, \citenamefont {Lucovsky}, \citenamefont {Geballe}, \citenamefont
  {Blank},\ and\ \citenamefont {Beasley}}]{siemons2007}%
  \BibitemOpen
  \bibfield  {author} {\bibinfo {author} {\bibfnamefont {W.}~\bibnamefont
  {Siemons}}, \bibinfo {author} {\bibfnamefont {G.}~\bibnamefont {Koster}},
  \bibinfo {author} {\bibfnamefont {H.}~\bibnamefont {Yamamoto}}, \bibinfo
  {author} {\bibfnamefont {W.~A.}\ \bibnamefont {Harrison}}, \bibinfo {author}
  {\bibfnamefont {G.}~\bibnamefont {Lucovsky}}, \bibinfo {author}
  {\bibfnamefont {T.~H.}\ \bibnamefont {Geballe}}, \bibinfo {author}
  {\bibfnamefont {D.~H.~A.}\ \bibnamefont {Blank}}, \ and\ \bibinfo {author}
  {\bibfnamefont {M.~R.}\ \bibnamefont {Beasley}},\ }\href@noop {} {\bibfield
  {journal} {\bibinfo  {journal} {Phys. Rev. Lett.}\ }\textbf {\bibinfo
  {volume} {98}},\ \bibinfo {pages} {196802} (\bibinfo {year}
  {2007})}\BibitemShut {NoStop}%
\bibitem [{\citenamefont {Zhong}\ \emph {et~al.}(2010)\citenamefont {Zhong},
  \citenamefont {Xu},\ and\ \citenamefont {Kelly}}]{Zhong2010}%
  \BibitemOpen
  \bibfield  {author} {\bibinfo {author} {\bibfnamefont {Z.}~\bibnamefont
  {Zhong}}, \bibinfo {author} {\bibfnamefont {P.~X.}\ \bibnamefont {Xu}}, \
  and\ \bibinfo {author} {\bibfnamefont {P.~J.}\ \bibnamefont {Kelly}},\
  }\href@noop {} {\bibfield  {journal} {\bibinfo  {journal} {Phys. Rev. B}\
  }\textbf {\bibinfo {volume} {82}},\ \bibinfo {pages} {165127} (\bibinfo
  {year} {2010})}\BibitemShut {NoStop}%
\bibitem [{\citenamefont {Chen}\ \emph {et~al.}(2011)\citenamefont {Chen},
  \citenamefont {Pryds}, \citenamefont {Kleibeuker}, \citenamefont {Koster},
  \citenamefont {Sun}, \citenamefont {Stamate}, \citenamefont {Shen},
  \citenamefont {Rijnders},\ and\ \citenamefont {Linderoth}}]{Chen2011_2}%
  \BibitemOpen
  \bibfield  {author} {\bibinfo {author} {\bibfnamefont {Y.}~\bibnamefont
  {Chen}}, \bibinfo {author} {\bibfnamefont {N.}~\bibnamefont {Pryds}},
  \bibinfo {author} {\bibfnamefont {J.~E.}\ \bibnamefont {Kleibeuker}},
  \bibinfo {author} {\bibfnamefont {G.}~\bibnamefont {Koster}}, \bibinfo
  {author} {\bibfnamefont {J.}~\bibnamefont {Sun}}, \bibinfo {author}
  {\bibfnamefont {E.}~\bibnamefont {Stamate}}, \bibinfo {author} {\bibfnamefont
  {B.}~\bibnamefont {Shen}}, \bibinfo {author} {\bibfnamefont {G.}~\bibnamefont
  {Rijnders}}, \ and\ \bibinfo {author} {\bibfnamefont {S.}~\bibnamefont
  {Linderoth}},\ }\href@noop {} {\bibfield  {journal} {\bibinfo  {journal}
  {Nano Lett.}\ }\textbf {\bibinfo {volume} {11}},\ \bibinfo {pages} {3774}
  (\bibinfo {year} {2011})}\BibitemShut {NoStop}%
\bibitem [{\citenamefont {Zaanen}\ \emph {et~al.}(1985)\citenamefont {Zaanen},
  \citenamefont {Sawatzky},\ and\ \citenamefont {Allen}}]{zaanen1985}%
  \BibitemOpen
  \bibfield  {author} {\bibinfo {author} {\bibfnamefont {J.}~\bibnamefont
  {Zaanen}}, \bibinfo {author} {\bibfnamefont {G.~A.}\ \bibnamefont
  {Sawatzky}}, \ and\ \bibinfo {author} {\bibfnamefont {J.~W.}\ \bibnamefont
  {Allen}},\ }\href@noop {} {\bibfield  {journal} {\bibinfo  {journal} {Phys.
  Rev. Lett.}\ }\textbf {\bibinfo {volume} {55}},\ \bibinfo {pages} {418}
  (\bibinfo {year} {1985})}\BibitemShut {NoStop}%
\bibitem [{\citenamefont {Arima}\ \emph {et~al.}(1993)\citenamefont {Arima},
  \citenamefont {Tokura},\ and\ \citenamefont {Torrance}}]{Arima1993}%
  \BibitemOpen
  \bibfield  {author} {\bibinfo {author} {\bibfnamefont {T.}~\bibnamefont
  {Arima}}, \bibinfo {author} {\bibfnamefont {Y.}~\bibnamefont {Tokura}}, \
  and\ \bibinfo {author} {\bibfnamefont {J.~B.}\ \bibnamefont {Torrance}},\
  }\href@noop {} {\bibfield  {journal} {\bibinfo  {journal} {Phys. Rev. B}\
  }\textbf {\bibinfo {volume} {48}},\ \bibinfo {pages} {17006} (\bibinfo {year}
  {1993})}\BibitemShut {NoStop}%
\bibitem [{\citenamefont {Chen}\ \emph {et~al.}(2013)\citenamefont {Chen},
  \citenamefont {Millis},\ and\ \citenamefont {Marianetti}}]{Chen2013}%
  \BibitemOpen
  \bibfield  {author} {\bibinfo {author} {\bibfnamefont {H.}~\bibnamefont
  {Chen}}, \bibinfo {author} {\bibfnamefont {A.~J.}\ \bibnamefont {Millis}}, \
  and\ \bibinfo {author} {\bibfnamefont {C.~A.}\ \bibnamefont {Marianetti}},\
  }\href@noop {} {\bibfield  {journal} {\bibinfo  {journal} {Phys. Rev. Lett.}\
  }\textbf {\bibinfo {volume} {111}},\ \bibinfo {pages} {116403} (\bibinfo
  {year} {2013})}\BibitemShut {NoStop}%
\bibitem [{\citenamefont {Blochl}(1994)}]{Blochl1994}%
  \BibitemOpen
  \bibfield  {author} {\bibinfo {author} {\bibfnamefont {P.~E.}\ \bibnamefont
  {Blochl}},\ }\href@noop {} {\bibfield  {journal} {\bibinfo  {journal} {Phys.
  Rev. B}\ }\textbf {\bibinfo {volume} {50}},\ \bibinfo {pages} {17953}
  (\bibinfo {year} {1994})}\BibitemShut {NoStop}%
\bibitem [{\citenamefont {Kresse}\ and\ \citenamefont
  {Joubert}(1999)}]{Kresse1999}%
  \BibitemOpen
  \bibfield  {author} {\bibinfo {author} {\bibfnamefont {G.}~\bibnamefont
  {Kresse}}\ and\ \bibinfo {author} {\bibfnamefont {D.}~\bibnamefont
  {Joubert}},\ }\href@noop {} {\bibfield  {journal} {\bibinfo  {journal} {Phys.
  Rev. B}\ }\textbf {\bibinfo {volume} {59}},\ \bibinfo {pages} {1758}
  (\bibinfo {year} {1999})}\BibitemShut {NoStop}%
\bibitem [{\citenamefont {Dudarev}\ \emph {et~al.}(1998)\citenamefont
  {Dudarev}, \citenamefont {Botton}, \citenamefont {Savrasov}, \citenamefont
  {Humphreys},\ and\ \citenamefont {Sutton}}]{Dudarev1998}%
  \BibitemOpen
  \bibfield  {author} {\bibinfo {author} {\bibfnamefont {S.~L.}\ \bibnamefont
  {Dudarev}}, \bibinfo {author} {\bibfnamefont {G.~A.}\ \bibnamefont {Botton}},
  \bibinfo {author} {\bibfnamefont {S.~Y.}\ \bibnamefont {Savrasov}}, \bibinfo
  {author} {\bibfnamefont {C.~J.}\ \bibnamefont {Humphreys}}, \ and\ \bibinfo
  {author} {\bibfnamefont {A.~P.}\ \bibnamefont {Sutton}},\ }\href@noop {}
  {\bibfield  {journal} {\bibinfo  {journal} {Phys. Rev. B}\ }\textbf {\bibinfo
  {volume} {57}},\ \bibinfo {pages} {1505} (\bibinfo {year}
  {1998})}\BibitemShut {NoStop}%
\bibitem [{\citenamefont {Tokura}\ \emph {et~al.}(1993)\citenamefont {Tokura},
  \citenamefont {Taguchi}, \citenamefont {Okada}, \citenamefont {Fujishima},
  \citenamefont {Arima}, \citenamefont {Kumagai},\ and\ \citenamefont
  {Iye}}]{tokura1993}%
  \BibitemOpen
  \bibfield  {author} {\bibinfo {author} {\bibfnamefont {Y.}~\bibnamefont
  {Tokura}}, \bibinfo {author} {\bibfnamefont {Y.}~\bibnamefont {Taguchi}},
  \bibinfo {author} {\bibfnamefont {Y.}~\bibnamefont {Okada}}, \bibinfo
  {author} {\bibfnamefont {Y.}~\bibnamefont {Fujishima}}, \bibinfo {author}
  {\bibfnamefont {T.}~\bibnamefont {Arima}}, \bibinfo {author} {\bibfnamefont
  {K.}~\bibnamefont {Kumagai}}, \ and\ \bibinfo {author} {\bibfnamefont
  {Y.}~\bibnamefont {Iye}},\ }\href@noop {} {\bibfield  {journal} {\bibinfo
  {journal} {Phys. Rev. Lett.}\ }\textbf {\bibinfo {volume} {70}},\ \bibinfo
  {pages} {2126} (\bibinfo {year} {1993})}\BibitemShut {NoStop}%
\bibitem [{\citenamefont {Koehler}\ \emph {et~al.}(1960)\citenamefont
  {Koehler}, \citenamefont {Wollan},\ and\ \citenamefont
  {Wilkinson}}]{Koehler1960}%
  \BibitemOpen
  \bibfield  {author} {\bibinfo {author} {\bibfnamefont {W.~C.}\ \bibnamefont
  {Koehler}}, \bibinfo {author} {\bibfnamefont {E.~O.}\ \bibnamefont {Wollan}},
  \ and\ \bibinfo {author} {\bibfnamefont {M.~K.}\ \bibnamefont {Wilkinson}},\
  }\href@noop {} {\bibfield  {journal} {\bibinfo  {journal} {Phys. Rev.}\
  }\textbf {\bibinfo {volume} {118}},\ \bibinfo {pages} {58} (\bibinfo {year}
  {1960})}\BibitemShut {NoStop}%
\bibitem [{\citenamefont {Pavarini}\ \emph {et~al.}(2004)\citenamefont
  {Pavarini}, \citenamefont {Biermann}, \citenamefont {Poteryaev},
  \citenamefont {Lichtenstein}, \citenamefont {Georges},\ and\ \citenamefont
  {Andersen}}]{Pavarini2004}%
  \BibitemOpen
  \bibfield  {author} {\bibinfo {author} {\bibfnamefont {E.}~\bibnamefont
  {Pavarini}}, \bibinfo {author} {\bibfnamefont {S.}~\bibnamefont {Biermann}},
  \bibinfo {author} {\bibfnamefont {A.}~\bibnamefont {Poteryaev}}, \bibinfo
  {author} {\bibfnamefont {A.~I.}\ \bibnamefont {Lichtenstein}}, \bibinfo
  {author} {\bibfnamefont {A.}~\bibnamefont {Georges}}, \ and\ \bibinfo
  {author} {\bibfnamefont {O.~K.}\ \bibnamefont {Andersen}},\ }\href@noop {}
  {\bibfield  {journal} {\bibinfo  {journal} {Phys. Rev. Lett.}\ }\textbf
  {\bibinfo {volume} {92}},\ \bibinfo {pages} {176403} (\bibinfo {year}
  {2004})}\BibitemShut {NoStop}%
\bibitem [{\citenamefont {Zhong}\ and\ \citenamefont
  {Kelly}(2008)}]{Zhong2008}%
  \BibitemOpen
  \bibfield  {author} {\bibinfo {author} {\bibfnamefont {Z.}~\bibnamefont
  {Zhong}}\ and\ \bibinfo {author} {\bibfnamefont {P.~J.}\ \bibnamefont
  {Kelly}},\ }\href@noop {} {\bibfield  {journal} {\bibinfo  {journal} {Eur.
  Phys. Lett.}\ }\textbf {\bibinfo {volume} {84}},\ \bibinfo {pages} {27001}
  (\bibinfo {year} {2008})}\BibitemShut {NoStop}%
\bibitem [{Sup()}]{SupplMat}%
  \BibitemOpen
  \href@noop {} {}\bibinfo {howpublished} {See for more details Supplemental
  Materials}\BibitemShut {NoStop}%
\bibitem [{\citenamefont {Zubko}\ \emph {et~al.}(2011)\citenamefont {Zubko},
  \citenamefont {Gariglio}, \citenamefont {Gabay}, \citenamefont {Ghosez},\
  and\ \citenamefont {Triscone}}]{Zubko2011}%
  \BibitemOpen
  \bibfield  {author} {\bibinfo {author} {\bibfnamefont {P.}~\bibnamefont
  {Zubko}}, \bibinfo {author} {\bibfnamefont {S.}~\bibnamefont {Gariglio}},
  \bibinfo {author} {\bibfnamefont {M.}~\bibnamefont {Gabay}}, \bibinfo
  {author} {\bibfnamefont {P.}~\bibnamefont {Ghosez}}, \ and\ \bibinfo {author}
  {\bibfnamefont {J.-M.}\ \bibnamefont {Triscone}},\ }\href@noop {} {\bibfield
  {journal} {\bibinfo  {journal} {Annu. Rev. Conden. Ma. P.}\ }\textbf
  {\bibinfo {volume} {2}},\ \bibinfo {pages} {141} (\bibinfo {year}
  {2011})}\BibitemShut {NoStop}%
\bibitem [{\citenamefont {Koster}\ \emph {et~al.}(1998)\citenamefont {Koster},
  \citenamefont {Kropman}, \citenamefont {Rijnders}, \citenamefont {Blank},\
  and\ \citenamefont {Rogalla}}]{koster1998}%
  \BibitemOpen
  \bibfield  {author} {\bibinfo {author} {\bibfnamefont {G.}~\bibnamefont
  {Koster}}, \bibinfo {author} {\bibfnamefont {B.~L.}\ \bibnamefont {Kropman}},
  \bibinfo {author} {\bibfnamefont {G.~J. H.~M.}\ \bibnamefont {Rijnders}},
  \bibinfo {author} {\bibfnamefont {D.~H.~A.}\ \bibnamefont {Blank}}, \ and\
  \bibinfo {author} {\bibfnamefont {H.}~\bibnamefont {Rogalla}},\ }\href@noop
  {} {\bibfield  {journal} {\bibinfo  {journal} {Appl. Phys. Lett.}\ }\textbf
  {\bibinfo {volume} {73}},\ \bibinfo {pages} {2920} (\bibinfo {year}
  {1998})}\BibitemShut {NoStop}%
\bibitem [{\citenamefont {Ohtomo}\ \emph
  {et~al.}(2002{\natexlab{b}})\citenamefont {Ohtomo}, \citenamefont {Muller},
  \citenamefont {Grazul},\ and\ \citenamefont {Hwang}}]{ohtomo2002}%
  \BibitemOpen
  \bibfield  {author} {\bibinfo {author} {\bibfnamefont {A.}~\bibnamefont
  {Ohtomo}}, \bibinfo {author} {\bibfnamefont {D.~A.}\ \bibnamefont {Muller}},
  \bibinfo {author} {\bibfnamefont {J.~L.}\ \bibnamefont {Grazul}}, \ and\
  \bibinfo {author} {\bibfnamefont {H.~Y.}\ \bibnamefont {Hwang}},\ }\href@noop
  {} {\bibfield  {journal} {\bibinfo  {journal} {Appl. Phys. Lett.}\ }\textbf
  {\bibinfo {volume} {80}},\ \bibinfo {pages} {3922} (\bibinfo {year}
  {2002}{\natexlab{b}})}\BibitemShut {NoStop}%
\bibitem [{Note1()}]{Note1}%
  \BibitemOpen
  \bibinfo {note} {Note that some Ti/Fe intermixing across the interface may be
  present, taking the low oxygen pressure during growth into account~\cite
  {Willmott2007}.}\BibitemShut {Stop}%
\bibitem [{Note2()}]{Note2}%
  \BibitemOpen
  \bibinfo {note} {No charging of the samples was observed during X-ray
  exposure since the SrTiO$_{3-\delta }$ became conducting as a result of the
  low oxygen pressure during growth and cool down.}\BibitemShut {Stop}%
\bibitem [{Note3()}]{Note3}%
  \BibitemOpen
  \bibinfo {note} {The La MNN (at $\sim $740-800~\protect \textit {e}V)
  obscures the Fe 2\protect \textit {p} satellite structure at higher binding
  energy. To allow proper normalization, we limited the Fe~2\protect \textit
  {p} range up to this satellite peak.}\BibitemShut {Stop}%
\bibitem [{Note4()}]{Note4}%
  \BibitemOpen
  \bibinfo {note} {\label {note1}Normalization of the valence band spectra is
  complicated by the Ti-O~2\protect \textit {p} and Fe-O~2\protect \textit {p}
  hybridization. To allow for a qualitative analysis, the valence band spectra
  were aligned on the intensity of the O~2\protect \textit {p} at 5~\protect
  \textit {e}V. However, this may result in minor normalization
  artefacts.}\BibitemShut {Stop}%
\bibitem [{\citenamefont {Fujii}\ \emph {et~al.}(1999)\citenamefont {Fujii},
  \citenamefont {de~Groot}, \citenamefont {Sawatzky}, \citenamefont {Voogt},
  \citenamefont {Hibma},\ and\ \citenamefont {Okada}}]{Fujii1999}%
  \BibitemOpen
  \bibfield  {author} {\bibinfo {author} {\bibfnamefont {T.}~\bibnamefont
  {Fujii}}, \bibinfo {author} {\bibfnamefont {F.~M.~F.}\ \bibnamefont
  {de~Groot}}, \bibinfo {author} {\bibfnamefont {G.~A.}\ \bibnamefont
  {Sawatzky}}, \bibinfo {author} {\bibfnamefont {F.~C.}\ \bibnamefont {Voogt}},
  \bibinfo {author} {\bibfnamefont {T.}~\bibnamefont {Hibma}}, \ and\ \bibinfo
  {author} {\bibfnamefont {K.}~\bibnamefont {Okada}},\ }\href@noop {}
  {\bibfield  {journal} {\bibinfo  {journal} {Phys. Rev. B}\ }\textbf {\bibinfo
  {volume} {59}},\ \bibinfo {pages} {3195} (\bibinfo {year}
  {1999})}\BibitemShut {NoStop}%
\bibitem [{\citenamefont {Kareev}\ \emph {et~al.}(2013)\citenamefont {Kareev},
  \citenamefont {Cao}, \citenamefont {Liu}, \citenamefont {Middey},
  \citenamefont {Meyers},\ and\ \citenamefont {Chakhalian}}]{Kareev2013}%
  \BibitemOpen
  \bibfield  {author} {\bibinfo {author} {\bibfnamefont {M.}~\bibnamefont
  {Kareev}}, \bibinfo {author} {\bibfnamefont {Y.}~\bibnamefont {Cao}},
  \bibinfo {author} {\bibfnamefont {X.}~\bibnamefont {Liu}}, \bibinfo {author}
  {\bibfnamefont {S.}~\bibnamefont {Middey}}, \bibinfo {author} {\bibfnamefont
  {D.}~\bibnamefont {Meyers}}, \ and\ \bibinfo {author} {\bibfnamefont
  {J.}~\bibnamefont {Chakhalian}},\ }\href@noop {} {\bibfield  {journal}
  {\bibinfo  {journal} {Appl. Phys. Lett.}\ }\textbf {\bibinfo {volume}
  {103}},\ \bibinfo {pages} {231605} (\bibinfo {year} {2013})}\BibitemShut
  {NoStop}%
\bibitem [{NIS()}]{NIST}%
  \BibitemOpen
  \href@noop {} {}\bibinfo {howpublished} {NIST standard reference database 71,
  version 2.1}\BibitemShut {NoStop}%
\bibitem [{\citenamefont {Takizawa}\ \emph {et~al.}(2006)\citenamefont
  {Takizawa}, \citenamefont {Wadati}, \citenamefont {Tanaka}, \citenamefont
  {Hashimoto}, \citenamefont {Yoshida}, \citenamefont {Fujimori}, \citenamefont
  {Chikamatsu}, \citenamefont {Kumigashira}, \citenamefont {Oshima},
  \citenamefont {Shibuya}, \citenamefont {Mihara}, \citenamefont {Ohnishi},
  \citenamefont {Lippmaa}, \citenamefont {Kawasaki}, \citenamefont {Koinuma},
  \citenamefont {Okamoto},\ and\ \citenamefont {Millis}}]{Takizawa2006}%
  \BibitemOpen
  \bibfield  {author} {\bibinfo {author} {\bibfnamefont {M.}~\bibnamefont
  {Takizawa}}, \bibinfo {author} {\bibfnamefont {H.}~\bibnamefont {Wadati}},
  \bibinfo {author} {\bibfnamefont {K.}~\bibnamefont {Tanaka}}, \bibinfo
  {author} {\bibfnamefont {M.}~\bibnamefont {Hashimoto}}, \bibinfo {author}
  {\bibfnamefont {T.}~\bibnamefont {Yoshida}}, \bibinfo {author} {\bibfnamefont
  {A.}~\bibnamefont {Fujimori}}, \bibinfo {author} {\bibfnamefont
  {A.}~\bibnamefont {Chikamatsu}}, \bibinfo {author} {\bibfnamefont
  {H.}~\bibnamefont {Kumigashira}}, \bibinfo {author} {\bibfnamefont
  {M.}~\bibnamefont {Oshima}}, \bibinfo {author} {\bibfnamefont
  {K.}~\bibnamefont {Shibuya}}, \bibinfo {author} {\bibfnamefont
  {T.}~\bibnamefont {Mihara}}, \bibinfo {author} {\bibfnamefont
  {T.}~\bibnamefont {Ohnishi}}, \bibinfo {author} {\bibfnamefont
  {M.}~\bibnamefont {Lippmaa}}, \bibinfo {author} {\bibfnamefont
  {M.}~\bibnamefont {Kawasaki}}, \bibinfo {author} {\bibfnamefont
  {H.}~\bibnamefont {Koinuma}}, \bibinfo {author} {\bibfnamefont
  {S.}~\bibnamefont {Okamoto}}, \ and\ \bibinfo {author} {\bibfnamefont
  {A.~J.}\ \bibnamefont {Millis}},\ }\href@noop {} {\bibfield  {journal}
  {\bibinfo  {journal} {Phys. Rev. Lett.}\ }\textbf {\bibinfo {volume} {97}},\
  \bibinfo {pages} {057601} (\bibinfo {year} {2006})}\BibitemShut {NoStop}%
\bibitem [{\citenamefont {Willmott}\ \emph {et~al.}(2007)\citenamefont
  {Willmott}, \citenamefont {Herger}, \citenamefont {Schlep\"utz}, \citenamefont
  {Martoccia}, \citenamefont {Patterson}, \citenamefont {Delley}, \citenamefont
  {Clarke}, \citenamefont {Kumah}, \citenamefont {Cionca},\ and\ \citenamefont
  {Yacoby}}]{Willmott2007}%
  \BibitemOpen
  \bibfield  {author} {\bibinfo {author} {\bibfnamefont {P.~R.}\ \bibnamefont
  {Willmott}, \bibfnamefont {S.~A.}\ \bibfnamefont {Pauli}}, \bibinfo {author} {\bibfnamefont
  {R.}~\bibnamefont {Herger}}, \bibinfo {author} {\bibfnamefont {C.~M.}\
  \bibnamefont {Schlep\"utz}}, \bibinfo {author} {\bibfnamefont {D.}~\bibnamefont
  {Martoccia}}, \bibinfo {author} {\bibfnamefont {B.~D.}\ \bibnamefont
  {Patterson}}, \bibinfo {author} {\bibfnamefont {B.}~\bibnamefont {Delley}},
  \bibinfo {author} {\bibfnamefont {R.}~\bibnamefont {Clarke}}, \bibinfo
  {author} {\bibfnamefont {D.}~\bibnamefont {Kumah}}, \bibinfo {author}
  {\bibfnamefont {C.}~\bibnamefont {Cionca}}, \ and\ \bibinfo {author}
  {\bibfnamefont {Y.}~\bibnamefont {Yacoby}},\ }\href@noop {} {\bibfield
  {journal} {\bibinfo  {journal} {Phys. Rev. Lett.}\ }\textbf {\bibinfo {volume}
  {99}},\ \bibinfo {pages} {155502} (\bibinfo {year} {2007})}\BibitemShut
  {NoStop}%
\end{thebibliography}
\end {document}